# Ferroelectricity, SSFLC, bistability and all that.

Ingolf Dahl

*MINA, Chalmers University, S-412 96 Göteborg, Sweden, email ingolf.dahl@telia.com*

**Abstract: In the book "Ferroelectric and Antiferroelectric Liquid Crystals" by S. T. Lagerwall, the concept "polar liquid crystals" is proposed for the concept earlier known as "ferroelectric liquid crystals", reserving the word "ferroelectric liquid crystals" for the case of "surface stabilization". Thus Lagerwall in this way, by redefinition, becomes the coinventor of "ferroelectric liquid crystals". The trouble is that a closer look on the invention reveals a state of bad logic and a total confusion. The concepts "polar", "ferroelectric", "hysteresis", "SSFLC" and "bistability" are essential in the writing of Lagerwall, but these words are not used in a rigorous way. Also the pictorial evidence used by Lagerwall to illustrate the discovery of surface stabilized liquid crystals raises several questions. An alternative view of the physics of SSFLC cells is presented.**

## Introduction

This is a review of the book "Ferroelectric and Antiferroelectric Liquid Crystals" by S.T. Lagerwall(Lagerwall 1999). In this context, it is necessary to review also the earlier work of Lagerwall especially the discovery of ferroelectricity in the chiral smectic C phase, as it is presented in the changing perspective of S.T. Lagerwall. I have during several years tried to discuss and criticise the views of Lagerwall internally, but it has been impossible to get a constructive discussion, and Lagerwall has exhorted me to discuss these matters in public. Lagerwall also consider himself a suitable person to discuss ethical matters and to criticise and report other researchers for scientific fraud, but the book and the research by Lagerwall touches some ethical questions, to be illustrated here. Lagerwall has also been a leading person in the field of ferroelectric liquid crystals, and his review papers have focused the research on some scientific questions, and defocused or confused some other questions in a questionable way.

Some of the pictures from the early work on ferroelectric liquid crystals are reprinted again and again. Since pictures have heavy impact on the thinking of people, it is important to point out that these pictures have a questionable origin and may not always show what is implicated.

A large part of the book is a word-by-word (and error-by-error) copy of the submission of Lagerwall to the Handbook of Liquid Crystals, edited by Demus(Demus, Goodby et al. 1998). Some parts are also copied word-by-word from an earlier paper (Lagerwall 1996).

Lagerwall finds it important to discuss some concepts in his book, namely "ferroelectric", "polar", "surface stabilization" etc., and he writes:

"*Coherent and contradiction-free terminology is certainly important, because vagueness and ambiguity are obstacles for clear thinking and comprehension.*" (Lagerwall 1999) (p. 4)

Lagerwall wants to give a very formal definition of the concept "ferroelectric", in order to implicate that he was involved in the discovery of ferroelectric liquid crystals. However, he succeeds to use this and related words in a vague and ambiguous way.

We must differentiate between common use of language and formal definitions. Common language is not entirely logical and consequent, and words are shifting in meaning dependent on the context. Formal definitions are important in mathematics and in legislation, but in physics the meaning of words is not so important, as long as we give the correct mathematical description. Discussion about the exact meaning of words often leads to unproductive hair splitting.

In this paper, I have made extensive citations from the writing of Lagerwall, to illustrate inconsistencies in the way of arguing.

**The discovery by Robert Meyer**

In the beginning, R.B. Meyer was recognized as the discoverer of ferroelectric liquid crystals:

*"As was discovered by R.B. Meyer, any tilted smectic phase built up by chiral molecules ought to have an intrinsic ferroelectric property in the sense that every smectic layer possesses an electric dipole density, **P**, which is perpendicular to the molecular tilt direction, **n**, and parallel to the smectic layer plane."* (Clark and Lagerwall 1980c).

*"Since 1974, when the first ferroelectric liquid crystals were reported by Meyer et al., the great potentialities of these new phases have been generally recognized, and extensive investigations on their electric properties and electro-optic response have been performed."…" All tilted smectics (C, F, G, H, I …. ) which are in addition chiral, are ferroelectric.* "(Clark, Handschy et al. 1983). Below a star "*" on a phase letter denotes the chiral variant of the phase.

In 1994 Lagerwall has changed his mind about the discovery of ferroelectricity in liquid crystals(Lagerwall and Stebler 1994):

*"The materials described in reference(Meyer, Liébert et al. 1975) were the first polar liquid crystals to be investigated."*

The 1996 paper: *"It was the Harvard physicist R.B. Meyer who first recognized that the symmetry properties of a chiral tilted smectic would allow a spontaneous polarization directed perpendicular to the tilt plane, and in collaboration with French chemists he synthesized the first such material in 1974. These were the first polar liquid crystals and as such something strikingly new. Of course, substances showing a smectic C\* phase had been synthesized accidentally several times before by other groups, but their very special polar character had never been detected. Meyer called these liquid crystals ferroelectric. As it subsequently turned out, this was almost but not quite true. But a most important first step had been taken towards finding a ferroelectric behavior.* " (Lagerwall 1996)(p. 9155)

*"Indeed, when very thin samples with d » 1 µm are made with the appropriate boundary condition, spontaneous ferroelectric domains all of a sudden appear in the absence of any applied field (figure 11). This second step was realized five years after Meyer's first paper. By applying an external field we can now get one set of domains to grow at the cost of the other and reverse the whole process on reversing the field. There are two stable states and a symmetric bistability; the response is according to figure 4(b). We might therefore call this structure a surface-stabilized ferroelectric liquid crystal (SSFLC). The surface stabilization brings the C\* phase out of its natural crystallographic state and transfers macroscopic polarization to the bulk. "* Reference (Lagerwall 1996) (p. 9157), also (Lagerwall 1999) (p. 67) by cut-and paste]

The implication of this is that Clark-Lagerwall are responsible of the discovery of ferroelectric liquid crystals. Figure 11 mentioned here shows a sample, probably in the smectic I\* or J\* phase, more about that below. Figure 4(b) shows the idealized P-E-curve for a typical ferroelectric material, reproduced in the section below about ferroelectric hysteresis. To my knowledge, no measurements of the electric hysteresis curve were made at the initial experiments on SSFLC cells, and as far as I know, such measurements on bistable cells have never been made in the Chalmers Liquid Crystal Group. The first current measurements in our group made by me and published 1984, but those measurements were made on monostable cells.

What R.B. Meyer had discovered was not a ferroelectric liquid crystal, it was "truly polar" liquid crystals:

*"Meyer also recognized in 1974 that all chiral tilted smectics would be truly polar and the first example of this kind, the helielectric smectic C\*, was presented in 1975."* (Lagerwall 1999) (p. 12)

**Ferroelectricity**

The local spontaneous polarization in the chiral smectic C phase has two-dimensional freedom: it can be oriented in any direction parallel to the smectic layers. In this respect this liquid crystal phase is different from many solid-state ferroelectrics, where the spontaneous polarization locally only can choose between two (or a discrete number of) different directions in a flip-flop manner. Robert Meyer, who postulated this behaviour of chiral smectic C, and also participated in the experimental verification, has the moral right to refine the word "ferroelectric" to describe the phase in the new context. To be more specific, one could denote this type 2D-ferroelectricity, to differentiate it from flip-flop ferroelectricity. The "hallmarks" of flip-flop ferroelectricity: bistability and the existence of domains, will not be a characteristic of a 2D-ferroelectric phase. Instead we should get a continuum of states, and continuous deformations between different points. Robert Meyer was aware of this, and described very soon the physical behaviour of chiral smectic C in an essentially correct way.

The words "ferroelectric" and "polar" are adjectives, which should be attached as attributes to substantives. If we want to give a definition of these words, it is important which substantives they are associated with. Are "polar" and "ferroelectric" properties of a substance? Or of a phase? Or of a structure or a geometrical configuration? Or of a interface between two materials? Or of a device or a compound system?

What is a medium? "*A promising way of overcoming these difficulties*" (with conventional liquid crystal devices) "*is to use chiral smectic C liquid crystals in a carefully chosen geometry. These media are ferroelectric and thus permit a very direct action by the external field.*" (Clark and Lagerwall 1980c)

About the SSFLC device: "*The attribute ferroelectric means that its response to an external electric field of an SSFLC is not any longer linear around the origin (as in the helielectric or antiferroelectric case) but strongly non-linear, with a certain field threshold (figure 11). This means that the structure is bistable and can take one of two different states of* *opposite polarization in the absence of any external field. In other words, it has memory properties. Moreover, the switching from one state to the other on reversing the sign of the field, is now extremely fast, obeying the simple equation*

$$\tau = \frac{\gamma}{P \cdot E}$$

*where tau is a characteristic switching time, gamma a characteristic viscosity, **P** the polarization and **E** the electric field. The switching time is found to be in the lower microsecond range. "…""Ferroelectric" also means that there exists a macroscopic polarization in the absence of an external electric field, which is switchable between two stable states. (If the macroscopic polarization is not switchable the state is called pyroelectric.) It may be noted that no ferroelectricity has been found to exist so far in the bulk of any liquid crystalline phase. Thus the smectic C\* phase is not ferroelectric per se. Surface stabilization is required for the appearance of macroscopic polarization and discrete stable states."* (Lagerwall August 2001)

So the SSFLC <u>device</u>, as a macroscopic object, is analogous to the unit cell or microscopic domain in solid state ferroelectric <u>phases</u>?

*"A ferroelectric phase shows the peculiarity of two stable states. These states are polarized in opposite directions (±P) in absence of applied field (E=0). The property in a material to have two stable states is called bistability."* (Lagerwall August 2001)

(Here ferroelectricity is a property of a phase, while bistability is a property in a material. Logical?)

**Ferroelectric hysteresis**

*"Figure 11. Response of polar dielectrics (containing local permanent dipoles) to an applied electric field; from top to bottom: paraelectric, ferroelectric, antiferroelectric and helielectric (helical antiferroelectric)."* (Lagerwall August 2001)

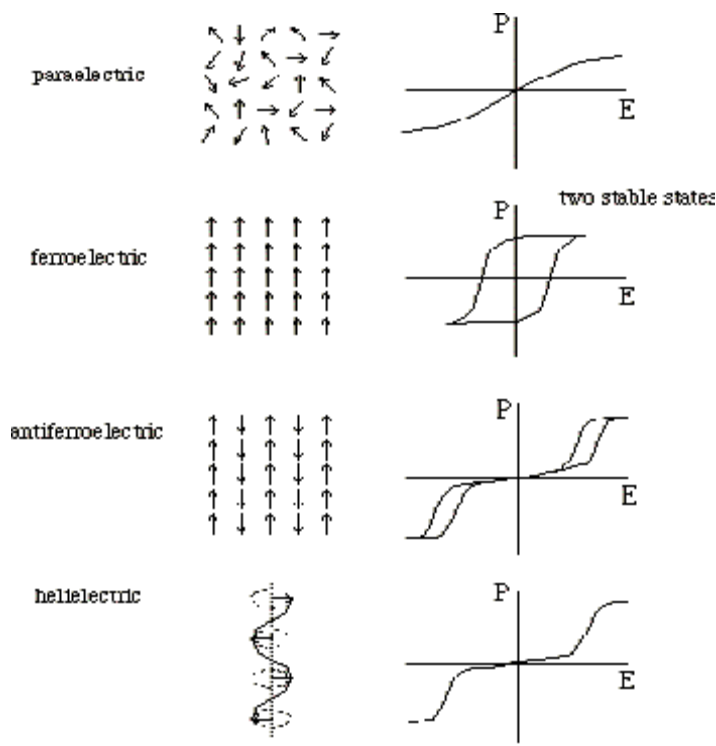

Fig. A: *"Figure 11. Response of polar dielectrics (containing local permanent dipoles) to an applied electric field; from top to bottom: paraelectric, ferroelectric, antiferroelectric and helielectric (helical antiferroelectric)."* (Lagerwall August 2001)

Definition of *"ferroelectric"*: *"There is a certain tendency to just call everything which has some polar character "ferroelectric". However understandable this might be, in particular because the word pyroelectric is not very descriptive when we are thinking of a polarization and not of a temperature effect, this habit should be strongly discouraged, as it only leads to confusion. Of course, the words piezo- and pyroelectric must also be used with the same care. It must be underlined that it is <u>the electric response</u> - cf figure 4 (the above figure) – which decides to which category a material belongs. Indeed, an important function of the name is to give information about the P-E relation."* (Lagerwall 1996)

*"…and it is also well known that certain loops (similar to ferroelectric hysteresis) may be obtained from a nonlinear lossy material…"* (Lagerwall 1999), (page 4) Oops! Then there must be some other, deeper definition of "ferroelectric".

Maybe, it could be a good idea to check the width of the hysteresis loop as function of the frequency of the applied voltage. If we have two bistable states in the material, with a potential barrier between, the width of the hysteresis loop should be independent of the frequency, while a nonlinear, lossy material should have a hysteresis loop width that is approx. proportional to the frequency. A non-linear, lossy material should thus switch if pulses of a characteristic area (voltage times pulse length) are applied.

*"Qualitatively, this makes it understandable why there is no threshold in the field as such, but rather in the voltage-time pulse area. Thus if we apply short pulses of 10 µs length, the required amplitude for switching might be, say 30 V, while, if we can afford pulses of 1 ms length (which we can in active of TFT driving) only a fraction of a volt might be necessary for switching between the two states. Without surfaces there would be no latching at all."* (Lagerwall 1999) (p. 309-310).

Oops again! Then the SSFLC might be a nonlinear, lossy device, instead of a ferroelectric!

*"It may be noted that no ferroelectricity has been found to exist so far in the bulk of any liquid crystalline phase. Thus the smectic C\* phase is not ferroelectric per se. Surface stabilization is required for the appearance of macroscopic polarization and discrete stable states."* (Lagerwall August 2001)

According to this macroscopic definition of "ferroelectric", a material to be tested should be inserted as a thin slice between two electrodes. Of course the experimentalist also is allowed to do almost anything to achieve a monodomain in the sample, as long as it remains in the same thermodynamic phase. With this definition, the property "ferroelectric" never refer to the "bulk" properties, it instead refer to the properties of the material as a thin slice. This is consistent with the properties of the polarization concept: it is not possible to define polarization without defining surfaces and boundaries of the system (explicit or implicit). So polarization is not a bulk property. It is also unphysical to discuss liquid crystals as "bulk materials", since there are long-range influences from the surfaces in almost all cases of theoretical or practical interest.

**The concept "polar"**

*"Finally, there seems to be a consensus about the concepts of polar and nonpolar liquids. Water is a polar liquid and mixes readily with other polar liquids, i.e. liquids consisting of polar molecules, like alcohol, at least as long as the sizes of the molecules are not too different, whereas it is insoluble in nonpolar liquids like benzene. If in liquid form, constituent polar molecules interact strongly with other polar molecules and, in particular, are easily oriented in external fields. We will also use this criterion for a liquid crystal. That is, we will call a liquid crystal polar if it contain local dipoles that are easily oriented in an applied electric field."* (Lagerwall 1999) (p. 9-10). Here evidently "polar" is the characteristic of a specific substance in a specific thermodynamic phase. (Water has relative dielectric constant 81, and ethanol has 26, while paraffin has 2.1. Commercial nematic liquid crystals have sometimes a relative dielectric constant of 20-30 parallel to the molecules.). Two pages later, Robert Meyer's discovery of "truly polar" liquid crystals is mentioned. So smectic C* is polar in the same sense as water? Or does the word "truly" implicate something else? Lagerwall has also written a whole article "Can liquids be macroscopically polar?" (Lagerwall 1996) without stating explicitly how he defines "macroscopically polar", and how the concept is related to the formal and common interpretations of the word "ferroelectric".

*"As we know today, nematics do not show ferroelectric or even polar properties."* (Lagerwall 1999) (p. 4)

From (Lagerwall August 2001):

*"polar: A polar molecule is a dipole. "*

*"Almost all liquid crystal molecules are dipoles, although the charge distribution normally is much more complicated than just described. A dipole in an electric field wants to turn such that the positive end points along the field and the negative one in the opposite direction. Therefore, one could be tempted to believe that this is the mechanism behind the electro-optic effects observed in liquid crystals. However, except in the very special cases of polar liquid crystals, which were introduced around 1975 by Robert B. Meyer and which since then are a very important field of liquid crystal research (see next page), this is not at all the case. In the majority of cases, in nematic and cholesteric materials, and in most smectic materials, the short range molecular organization is such that the local dipoles are everywhere compensating each other on the average, and this tendency to anti-parallel order is so strong that it is not at all affected by external electric fields. In other words, there is a "head-and-tail" symmetry in the distribution of the rodlike molecules. There are always as many molecules "head up" as "head down". The nematic liquid crystal therefore has quadrupolar order and not dipolar (polar) order, which is also the reason why we can change sign on the local director (**n** -> -**n**) without changing any macroscopic properties of the material.*

*The mechanism behind the fact that the local optic axis is affected by an external electric field is instead based on the anisotropic properties of the nematic; in this case the important point is that the dielectric constant is different in the direction along the director from the one in the direction perpendicular to it. Most often the dielectric constant is larger along **n**, and because the director turns in such a way that the maximum value of the dielectric constant lies along the direction of the field, the director, and hence the optic axis, orients along the field for such molecules.*

*In a solid the dipoles are too tightly bound to be easily reoriented by an electric field. In a normal liquid the thermal motion of the molecules normally overcomes the tendency for the*

*dipoles to orient. In a polar liquid crystal, on the other hand, we find just the right combination of order and flexibility to make the dipoles follow the sign of the field. "*

The tendency to anti-parallel ordering in the nematic phase sometimes lowers the average dielectric constant somewhat at the transition liquid-to-nematic-liquid-crystal. However, the change is mostly less than one unit in the relative dielectric constant (see Fig. 84 and 85 in the Lagerwall book (Lagerwall 1999)), and compared to the large dielectric constants of some nematics, this must be considered as a small effect. Thus, a nematic liquid crystal with polar molecules could easily be as polar as liquid alcohol, and we could affect the "head-tail symmetry" to approximately the same extent. But if we want to characterize the phases as such, not the substances, of course both the liquid phase and the nematic phase should be considered as non-polar. The term "a polar liquid crystal" is thus ambiguous.

"Polar" is not a good word for the characterisation of the properties of the chiral smectic C phase. The word has several meanings in common language, especially in scientific contexts (polar star, polar vectors, polar coordinates, polar surfaces, polar interaction etc.), and does not in itself implicate that the dipoles are easily reoriented. Consider, for example, the following definition, also by Lagerwall(!) (Lagerwall 1996) (p. 9147):

*"By polar materials we mean any kind of matter characterized by local dipoles, for instance a liquid consisting of dipolar molecules or a solid with dipoles distributed randomly or with any kind of specific order, for instance such that we have two sublattices with polarization that is homogeneous but opposite in direction, compensating each other (antiferroelectric)."*

The word "polar" is ambiguous when it is used both to describe qualitative and quantitative properties. That the substance water is a polar liquid is a quantitative statement. That the nematic phase is non-polar is a qualitative statement.

The word "helielectric" is sometimes used as a suitable replacement word for the characterisation of the chiral smectic C phase. That word is not without difficulties either. It is motivated by the fact that the so called "bulk ground state" of chiral smectic C is helically twisted. If we then would demand that our language should be logical and consequent, the ferromagnetic phase of magnetic materials should be renamed "domainmagnetic", because the aligned ferromagnetic state with its high magnetic energy mostly does not represent the lowest energy state… But a thermodynamic phase contains more than just the ground state.

**The SSFLC concept**

The SSFLC concept sometimes seems to include all cases, when the surface forces unwind the helix in thin, smectic C samples in bookshelf or quasi-bookshelf geometry. We can denote this "*the wide definition of SSFLC*". Sometimes specific boundary conditions are assumed ("*the narrow definition*"), sometimes introduced just as a possible example, but the conclusions from these assumptions are often generalized to be valid or "typical" for SSFLC devices according to the wide definition.

*"As stated above, the expression SSFLC itself was not coined until 1983 in reference (Clark, Handschy et al. 1983)."* (Lagerwall 1999) (p. 71)

From the paper(Clark, Handschy et al. 1983) (Note the use of the word "basic"! This distinction is soon forgotten.): *"In 1980, a new ferroelectric smectic device structure, the Surface Stabilized Ferroelectric Liquid Crystal (SSFLC) structure, was reported," showing that a very fast, bistable, electro-optic response with threshold is achievable in ferroelectric liquid crystals.",…,"In the SSFLC structure, surface interactions are used to unwind the spontaneous helix. The basic SSFLC structure" combines three appropriately chosen features (sample thickness, layer direction, boundary conditions), and is shown in Figure 1."* (Fig. B)

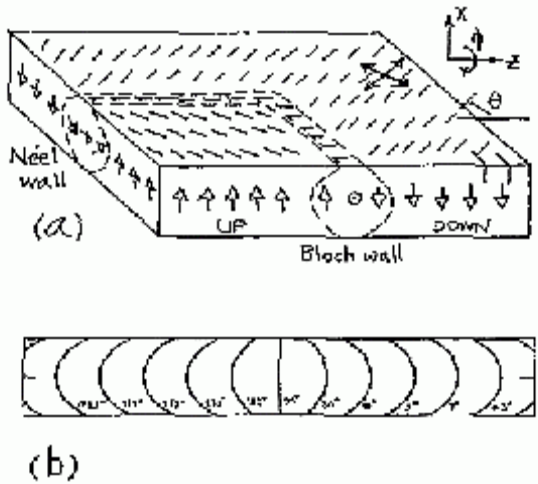

Fig. B: *"FIGURE I a) Schematic of the Surface Stabilized Ferroelectric Liquid Crystal Structure, showing the tilted smectic with layers normal to the bounding plates, the director orientation, **n** (short lines), and the accompanying ferroelectric polarization, **P** (arrows). The bounding plates are treated to constrain the director to be parallel to their surfaces, but free to reorient in the surface plane. The result is two stable states of the director field labeled here as UP and DOWN, according to the orientation of **P**. The displayed structure has both states present, separated by a domain wall, which in analogy to ferromagnets, is a Neél wall where it runs normal to the layers and a Bloch wall where it runs along the layers. b) Contours of constant orientation resulting from the continuum calculation of SSFLC domain wall structure presented in Section IV."*

"We now point out several unique and novel features of the SSFLC geometry. With the helix unwound, a ferroelectric smectic in the SSFLC geometry behaves optically essentially as a uniaxial slab with the uniaxis along the director orientation. The effect of switching is to rotate the uniaxis about the normal to the surface through an angle of twice the tilt angle 0. The SSFLC is the only liquid crystal parallel-plate geometry allowing a rotation of the uniaxis of a homogeneous sample about the surface normal. Another SSFLC feature to be noted is the nature of the required boundary condition. In order to obtain bistability, *boundary conditions which constrain the molecules to be parallel to the plates, but allow several or continuous orientations about the normal to the plates are required. The SSFLC is the first liquid crystal electro-optic structure to employ such a combination of strong-weak boundary conditions. A consequence of this feature and an essential SSFLC property is that the director at the surfaces is switched between stable surface orientation states as an essential part of the overall switching process. The SSFLC is the first liquid crystal electro-optic structure to successfully employ multistate surfaces and the only structure wherein switching between stable surface states has been demonstrated."*

The wide definition: *"With this layer orientation"* (not parallel to the interface) *"the interfaces which results establish orientational anisotropy within the layer planes. This breaks the degeneracy in $\varphi$"* (the azimuthal angle)*", and with sufficiently strong interaction and with sufficiently thin sample the helix will be destabilized to produce new interface-stabilized states of $\varphi$. Thus, the hallmark of the SSFLC device is the absence of the helix under static and dynamic states of $\varphi$ conditions in which it would be present in a bulk sample."* (Clark and Lagerwall 1991) (p. 34)

When I read a draft of the 1991 chapter, I commented "You also use the word 'hallmark' on the same page. This word is a bit frustrating, because it does not tell the reader if the properties stated as 'hallmarks' are defining properties, or if they are deduced." Since the word hallmark still is there in the final version, I must conclude that it left there intentionally to deliberately confuse the reader.

The wide definition is said to imply the narrow: *"It* (the helix) *can, however, also be suppressed by surfaces, in a special so-called bookshelf geometry (smectic layers perpendicular to the surfaces), when these come sufficiently close together, typically one or two micrometers. In this case spontaneous domains of UP and DOWN polarization appear (figure 10) and the material between the surfaces acquire ferroelectric properties (Clark and Lagerwall, 1980). Such a structure is therefore called a Surface-Stabilized Ferroelectric*

*Liquid Crystal (SSFLC)."* (Lagerwall August 2001), with Fig. 10 explicitly displaying a smectic F* cell!

Lagerwall beats the big drum when writing about SSFLC, using a special pompous style of writing:

*"Such cells can switch with speeds beyond the megahertz range. Furthermore, they have the intrinsic property of being bistable, by geometry, if the boundary conditions are carefully chosen."* (Clark, Lagerwall et al. 1985) There is a large difference between "within the megahertz range", 1 Mhz to 1 GHz, and "beyond the megahertz range", which implies 1 GHz or above, or at least 10 MHz or above. With ferroelectric liquid crystals, it is usually difficult to reach 1 MHz.

*"The surface-stabilized ferroelectric liquid crystal (SSFLC) is a structural concept which enables the basic smectic C\* energy degeneracy to be lifted. It employs the interaction of the **n-p** orientation field with the planar interfaces which are introduced when the liquid crystal is put between electroded plates in the typical electrooptic geometry as the dielectric in a transparent capacitor. The orientation anisotropy intrinsic to planar interfaces is transmitted to the volume of the liquid crystal by the surface interaction and the deformation energy of the **n-p** field, lifting the orientational degeneracy and making possible new stable or metastable **n-p** structures **which are not achievable in the bulk**, including the appearance of walls and ferroelectric domains. These new structures are intrinsic properties of the combination of characteristics of the liquid crystals with those of the interfaces. It is indeed these structures, including the helix-free domains and the motion of their boundaries under an applied field, that are the identifying hallmark of SSFLCs. SSFLCs exhibit not only unique structures but also electrooptic responses and switching properties which cannot be found in bulk phases. An appreciation of their richness and complexity has led to the gradual development of a new branch of liquid crystal physics devoted to their understanding. This work has to date uncovered a variety of novel effects and has added important new features to*

*the phenomenology of liquid crystals, including interfaces which are switchable between stable orientation states, liquid crystal-liquid crystal interfaces, first-order orientation transitions, and novel layer and defect structures. The many possible combinations of SSFLC surface interactions, layer geometries, and material properties have only begun to be explored, so that the field promises to be an active one"* (Clark and Lagerwall 1991) (p. 25). This citation does not support the view that the wide definition of SSFLC implies the narrow. Moreover, structures, which include surfaces in their definition (for instance walls), cannot by definition be achieved in the bulk. Helix-free volume regions are achievable in the bulk. The "domain" notation is a bit misleading, because magnetic domains are often enclosed by discontinuous walls, where the magnetization changes abruptly when we pass a two-dimensional surface. Discontinuous walls inside a smectic C*-cell require two-dimensional layer ordering defects.

**The pictures of bistable domains: SmC or SmF? DOBAMBC or HOBACPC?**

The two substances used in the initial experiments were DOBAMBC and HOBACPC. According to Robert Meyer *et al.* (Meyer, Liébert et al. 1975) DOBAMBC has the phase sequence for falling temperature:

Isotropic (117º) Smectic A (95º) Smectic C* (63º) Smectic H ( -) Crystal.

On heating we get

Crystal (76º) Smectic C*.

The smectic H phase has later been redefined as smectic I* by Goodby(Guillon, Stamatoff et al. 1982).

According to Keller *et al* (Keller, Jugé et al. 1976), the phase sequence of HOBACPC is

Isotropic (136.5º) Smectic A (81º) Smectic C* (74.5º) Smectic H (65º) Cryst al.

According to Martinot-Lagarde et al. (Martinot-Lagarde, Duke et al. 1981) the phase sequence is

Isotropic (135º) Smectic A (78º) Smectic C* (64º) Smectic H (60º) Crystal.

According to Jain and Wahl(Jain and Wahl 1983), the phase sequence is

Isotropic (135.8º) Smectic A (80º) Smectic C* (73.4º) Smectic I* (67.5º) Smectic G´ or Smectic H´ (60º) Crystal.

A similar phase sequence was already proposed by Doucet(Doucet, Keller et al. 1978). The substance has poor stability, which especially might lower and shrink the C phase. We see here that the former identified C* phase has been split into two phases, C* and I*, which indicates that old measurements on HOBACPC must be reinterpreted. The nature of the 72º measurements is unclear, and the 68º measurements most probably were made in the I* phase. There is no such reinterpretation anywhere what I have found. The old measurements are instead used as a support for the description of the behavior of smectic C*:

*"We may recall that the submicrosecond switching reported in the first paper had been observed on a single substance, HOBACPC, at an elevated temperature of 68 ºC. "* (Lagerwall 1999) (p. 215)

In the paper (Clark and Lagerwall 1981) the phase sequence seems to be that proposed by Martinot-Lagarde, with the exception that the H phase instead was denoted as the F phase. In 1984, Lagerwall(Lagerwall and Dahl 1984), in section 5, had become aware of the new phase sequence proposed by Jain and Wahl, and identified the lowest smectic phase in HOBACPC as smectic J*, instead of Smectic G´ or Smectic H´. In the same paper, section 7, the I-phase notation is said to replace the F-phase notation, which Clark and Lagerwall had used for the lowest smectic phase, below 64º. The exact phase sequence is thus not important to Lagerwall.

In one of their first reports(Clark and Lagerwall 1980c) Clark and Lagerwall report on pulse switching of HOBACPC at 88º, probably in the smectic C phase. (It should be in the smectic-A phase if the temperatures reported by Keller and Martinot-Lagarde were correct.) In the paper (Clark, Handschy et al. 1983), Clark, Handschy and Lagerwall also report, with a diagram (Fig. 6a and c), on pulse switching of HOBACPC at 72º and 68º in the smectic-C phase. The pulse lengths required for the 88º and the 72º measurements are very similar, while the 68º switching requires longer pulses. Was the temperature 88º correct? And was the sample really in the C-phase at 68º?

These figures present more questions. Fig 6a in (Clark, Handschy et al. 1983) seem to be redrawn from Fig 3a in (Clark and Lagerwall 1980a), with reversed sign of optic response. However, the thickness reported in (Clark and Lagerwall 1980a) is 1.5 micrometers and the sampled area is 200 µm times 200 µm (40 000 $m$m$^2$). In (Clark, Handschy et al. 1983) the thickness is 1 µm and the area 200 µm$^2$. Figure 6c in (Clark, Handschy et al. 1983), earlier also published in (Clark and Lagerwall 1981) as Fig 6, is remeasured and redrawn in (Clark, Lagerwall et al. 1985) with more details and new temperatures. see Fig. C. The base temperature $T_c$ is 80º, and the two intermediate new curves are said to belong to the I* phase, at 73º and 69º, while the uppermost curve represent the J* phase, then at 66º.

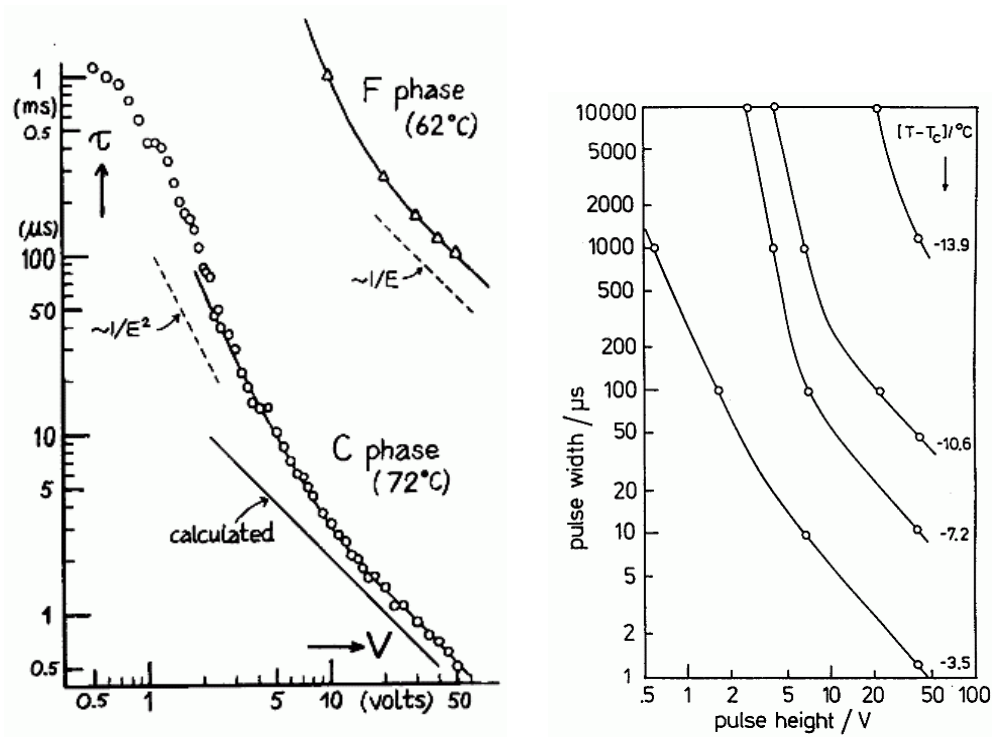

Fig. C: Relation between pulse time and required switching voltage, from (Clark and Lagerwall 1981) and from (Clark, Lagerwall et al. 1985)

In the report(Clark and Lagerwall 1980c) the exact phase of the liquid crystal is mentioned only twice. In the introduction, the promising features of fast and bistable, smectic C electro-optic devices are presented (wishful thinking?). However, the pictures presented are said to show bistability in a DOBAMBC sample in the smectic H phase. Did "microsecond-speed, bistable" mean fast switching in the C phase and bistable switching in the F or I phase?

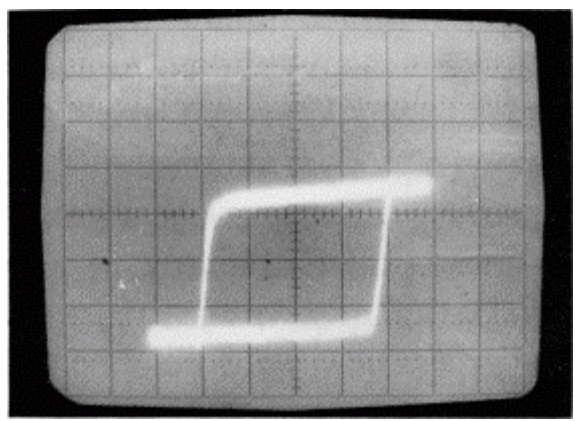

Fig D: Optical hysteresis loop

In the paper from 1981(Clark and Lagerwall 1981) an optical hysteresis loop is shown in figure 5, displaying the photodiode response of triangular voltage (4 volts peak to peak, 0.1 Hz) "demonstrating threshold, saturation, memory and symmetric bistability" in the smectic-F phase of DOBAMBC at 60º.

In the paper (Clark, Handschy et al. 1983), the same picture (upside-down) is shown in figure 6d. The figure text implies that it is measured in smectic C* HOBACPC at 68º, with applied triangle wave with 3 volts peak to peak. The main text instead tells that this measurement is done on a DOBAMBC sample in the smectic F* phase, 4 volts peak to peak.

In 1991(Clark and Lagerwall 1991), the figure text talks about smectic C* HOBACPC at 60º with approx 10 volts peak to peak.

This figure is also reproduced in the Lagerwall book (Lagerwall 1999) (figure 21). In the figure text it is reported as DOBAMBC at 60º, which implies I* phase. Now instead the main text uses it as a motivation for ferroelectricity in the C* phase.

Confusing? What the optical hysteresis loop really displays is probably the movement of disclination lines or domain walls over a laser spot in the microscope(Clark and Lagerwall 1980a). This movement is viscously controlled, see the picture sequence discussed later in this paper. Maybe it is this optical hysteresis loop that has inspired Lagerwall-Stebler to the following statements about the electrical hysteresis loop:

*"Second, the relation between polarization P and applied field has been completely changed, as shown in Figure 8, to a hysteresis loop which is found even at such low frequencies that one can exclude spurious viscous effects as sometimes seen in dielectrics. There is now a non-zero coercive force and remanence; a non-zero polarization in the absence of an electric field. Furthermore, the hysteresis curve is generally quite symmetric, indicating that the two polarization states of opposite sign are both stable. These properties are the characteristics of ferroelectricity."* (Lagerwall and Stebler 1994)

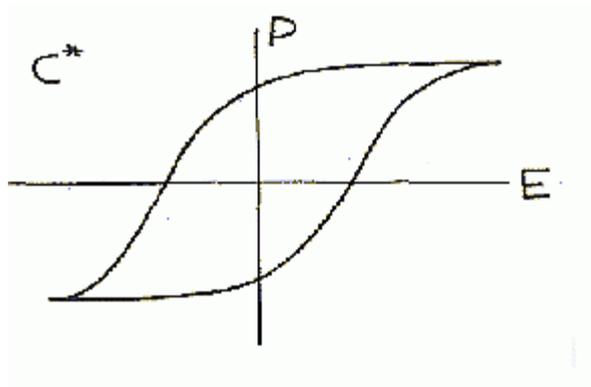

Fig. E: *"Figure 8. Relation between polarization P and electric field E for the surface-stabilized smectic C\* state, showing the existence of a spontaneous polarization in the absence of external field."* (Lagerwall and Stebler 1994)

Neither Lagerwall nor Stebler have to my knowledge done any electro-optical measurements on ferroelectric liquid crystals on their own. It is very difficult to make such measurements at "*even at such low frequencies that one can exclude spurious viscous effects"*, because any ion content in the cell will mask the polarization reversal current at low frequencies. At higher frequencies, the polarization reversal is always viscously controlled, and the width of the hysteresis curve depends on the frequency. The hysteresis loop will also show up in mono-stable samples for higher frequencies. The statement here about the width of the hysteresis curve is evidently in contradiction to the citation given above about threshold in the voltage-time pulse area.(Lagerwall 1999) (p. 309-310).

The photograph below appeared the first time in the paper by Lagerwall-Dahl, with reversed contrast. (Lagerwall and Dahl 1984), see Fig. F. I was coauthor of this paper, but contributed only to the sections 1, 2, and 4, and was not aware of the unclear phase of this picture.

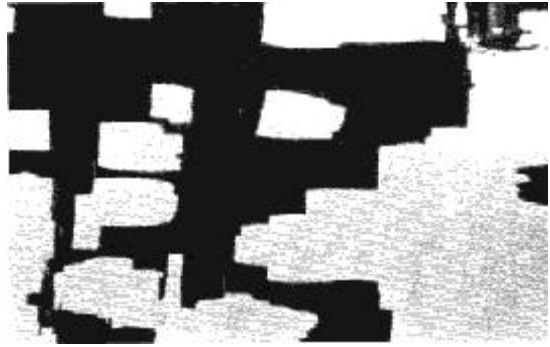

Fig. F: From (Lagerwall and Dahl 1984)

It has been given the following figure texts:

*"Figure 13. Ferroelectric domains with UP and DOWN polarization state appear when the spontaneous helical director structure is suppressed by surface action in the SSFLC structure. On the horizontal cross section the orientation of the optic axis (n field) is shown, on the vertical side sections the corresponding polarization (**P** field). Walls (of a mixed Bloch-Neel character in the **P** field) separate domains of opposite sign. By changing the polarizer setting the wall itself can be made visible rather than the domains (several examples of this are given in ref.55(Clark, Handschy et al. 1983)). The thickness of the wall is most often comparable to the cell thickness."* (Lagerwall and Dahl 1984). In the main text this picture is said to show a smectic C medium.

*"Figure 5 Spontaneous domains in a surface stabilized ferro-electric liquid crystal. The appearance of such domains (here is shown one of the first observation, from 1978, on the substance DOBAMBC) is a hallmark of SSFLC structures."* (Lagerwall, Clark et al. 1989) The smectic F* and I* phases are not mentioned in the paper! The domains here are said to be separated by surface stabilized walls. If the figure is from 1978, why were clearly inferior pictures published in the early papers?

*"Figure 4.2. Domains of opposite ferroelectric polarization in smectic F* HOBACPC. Image field is 320 micrometers. From (Clark, Handschy et al. 1983)."* (Clark and Lagerwall 1991). I have not found this picture in the paper (Clark, Handschy et al. 1983). I commented a draft to the paper (Clark and Lagerwall 1991) in the following way:

"The photos 12 (the above) and 13 (below) are said to be of smectic-C HOBACPC. Are you sure that they show smectic C? They very much resemble the video recordings (from August 1979) of the H phase, and the disclination lines between the differently oriented regions show a preference to be either parallel or normal to the smectic layers, and this I interpret as a characteristic of a lower smectic phase. See for instance Gray & Goodby(Gray and Goodby 1984), plate 35 or Demus & Richter(Demus and Richter 1980) plate 167 or 172. On the other hand, they also resemble Fig 77 (b), which should be a chevron sample, if I interpret the information properly."

Thus the figure text was corrected after my comment, but the photo above was still discussed in the text in the smectic-C* context. The smectic F*-phase was only discussed very shortly, together with other tilted smectic phases. In the book "the Physics of Liquid Crystals" by P.G. de Gennes and J. Prost (de Gennes and Prost 1993) the same figure appear again, with the figure text:

"Fig. 7.26. First reported picture of the ferroelectric surface-stabilized state (Courtesy S. Lagerwall). The dark domains correspond to a molecular direction parallel to the polarizer. Note the good contrast between the two states. The spontaneously appearing domains corresponds to a polarization pointing towards and away from the observer, respectively, and give the surface-stabilized state properties that are characteristic of conventional, solid state ferroelectrics. In particular, the memorized state can be reversed by an applied electric field."

In the main text, the figure is referred to in the following way:

*"7.2.7.4 Importance of the bookshelf geometry*

*This geometry provides a very promising display device(Clark and Lagerwall 1980a) which is characterized by good contrast, bistability, and speed.*

*Coupling to light. The first point is easily understood by observing Fig. 7.26: viewing in transmission between crossed polarizers under suitable conditions, a $S_C^*$ slab in the bookshelf geometry reveals black and white domains with a remarkable contrast."*

One can almost hear Lagerwall's pompous style of writing sound through. A peculiar and funny thing about this citation is that the reference to the 1980 paper(Clark and Lagerwall 1980a) is given erroneously in the book, with "I. Dahl" as co-author.

*"Figure 11. Spontaneous ferroelectric domains appearing in the surface-stabilized state."* (Lagerwall 1996) The smectic F*or I* phases are not mentioned in this paper! The text discusses the smectic C* phase, and comments this picture in the following way, not mentioning the required shear to align the cells: *"Indeed, when very thin samples with d » 1 **m**m are made with the appropriate boundary conditions, spontaneous ferroelectric domains all of a sudden appear in the absense of any applied field (figure 11)."*

*"Figure 10. Domains of opposite ferroelectric polarization in smectic F\* phase (a tilted smectic phase lying below the smectic C\* phase in temperature) in the substance HOBACPC. Image field width is 320 micrometers."* (Lagerwall August 2001)

*"Figure 17. Spontaneous ferroelectric domains appearing in the surface-stabilized state."* (Lagerwall 1999)]. The smectic F* or I* phases are not at all mentioned in the book! **Instead this photo is used, together with the optical hysteresis loop discussed above, as the prime indicators of the discovery of surface-stabilized ferroelectric liquid crystals (SSFLC) in the C\* phase!**

Another set of figures (Fig. G):

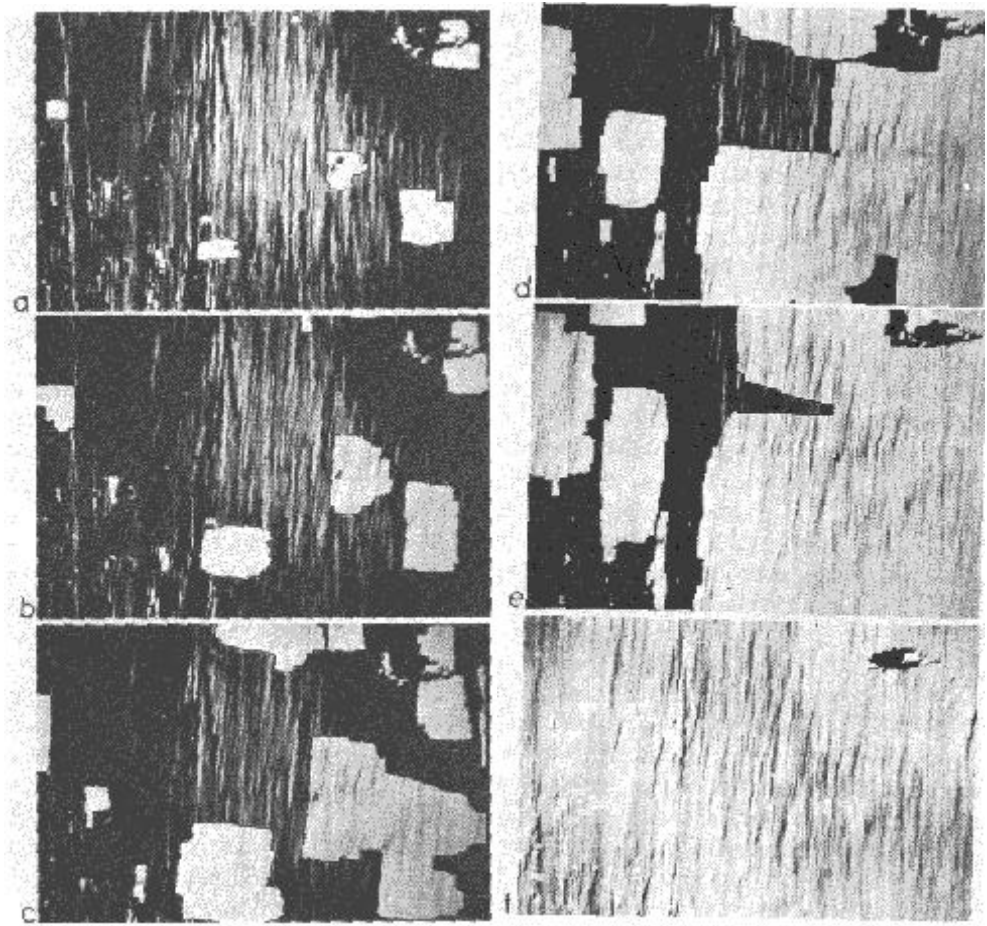

Fig G: *"FIGURE 5 A sequence of SSFLC domains in a 1.5 micrometer thick smectic C\* HOBACPC sample showing domain wall motion in the LOW field regime in response to a slowly increasing voltage of 0 V (a) to 1.5V (f). The moving domain walls encounter pinning sites which are successively overcome as E is increased. In e) the narrow extension of the dark domain to the right is caused by a pair of strong pinning sites impeding wall motion (image field width = 320 micrometers). "* (Clark, Handschy et al. 1983)

In the paper from 1991 (Clark and Lagerwall 1991) the same text as in the original is repeated, but the pictures are said to display smectic F* phase of HOBACPC, and the following text is added: *"The domains coalesce without forming disclinations because of the pretilt of **j**, and therefore its direction of change during reorientation is the same for all domains"*. However, the figures are discussed in a C* context. Some of the pictures are also replaced: A new figure *a* (black monodomain) is inserted. Then the figures *a*, *b*, *c*, *d*, and *e* from the paper(Clark, Handschy et al. 1983) follow, omitting figure *f*. We can conclude that the voltage for figure *d* should be either 1 V or 1.25 V.

In the paper(Lagerwall and Stebler 1994), figure *d* in the sequence above is discussed again in a C* context, with the following figure text (read carefully!):

*"Figure 9. Coexistence of up and down polarization states in adjacent domains (here in a rather symmetric fashion, showing the energetic equivalence of the two states) is a characteristic new phenomenon immediately revealing a surface-stabilized ferroelectric liquid crystal. The actual micrograph shows what may be described as the specimen's maiden state. It so far never experienced any selecting electric pulse, that could otherwise had given it altogether opaque or transmitting appearance."*

In the main text, this figure is discussed in the following way:

*"We will limit ourselves to a simplified discussion of some common cases, in which the surfaces would admit the up and down states of polarization in an essentially symmetric manner. It is far from trivial to realize such boundary conditions in practice, but it can be made. In this case spontaneous domains of opposite polarization states appear. These are ferroelectric domains. Because of the multiple use of the word "domain" in the context of liquid crystals and the different use of the word in the context of ferroelectrics it may be pointed out that the concept "ferroelectric domain" may only designate a region of non-zero polarization which appears spontaneously, i.e. in the* absence *of any applied field. The micrograph of Figure 9 shows the characteristic appearance of such domains representing up and down polarization, as seen between crossed polarizers."*

In one of early papers (Clark and Lagerwall 1980a), Clark and Lagerwall show fuzzy transmission micrographs of smectic C* DOBAMBC samples at 84º. One set of pictures, reminiscent of the pictures shown above, display domain wall migration without bistability: *"this observation prompted for the search for bistable operation"*. Why has this texture been reinterpreted? Another set, quite different in appearance and evidently from another area in the sample, is said to illustrate bistability. Can we really trust the authors that these really are smectic C* samples? And if the bistability is a natural consequence of the boundary conditions, why have they chosen such a messy area to illustrate the bistability feature? Did they study an area with chevron structure?

In the papers from 1989, 1994, and 1996 and in his book (Lagerwall, Clark et al. 1989; Lagerwall and Stebler 1994; Lagerwall 1996; Lagerwall 1999), Lagerwall does not even mention the existence or properties of the smectic I* and F* phases.

**The smectic F* and I* phases**

Is it important if the pictures are from smectic C*, smectic F*, or smectic I*? Yes, indeed. The smectic F and I phases are also tilted smectic phases, but they differ from the C phase in that there is a "hexatic bond ordering" present. In each smectic layer, the molecules form a usually imperfect lattice, with most of the molecule surrounded by six nearest neighbours. The lattices in adjacent smectic layers are not fixed in their relative position, but they are aligned directionally. Thus, the bonds between adjacent molecules are not isotropically distributed, they are instead aligned parallel to three directions. In orthogonal smectic phases, such bond ordering gives six-fold rotational symmetry around the layer normal. In the tilted smectic layers, the bond directions interact with the tilt direction. The difference between the I and the F phase is that the tilt direction in the I phase is along a bond direction, while it in the F phase is along a bisector of the bond directions. The six-fold rotational symmetry of the bond directions should then be slightly disturbed and reduced to twofold symmetry. The bond order stiffens then the tilt direction, which means that the conical degeneracy of the smectic C phase is replaced by local two-fold symmetry for fixed directions of the bond directions. Elastic distortions, without defects, will energetically be more costly than in the smectic C phase, however "paramorphic" textures can be found upon cooling from the C phase. Also the helical structure is sometimes present. The I* and F* phases are on the border line between 2D and flip-flop ferroelectricity: reorientation of the tilt 180 degrees does not require change of the bond directions, but a small angle rotation of the tilt direction require such a change. This implies that the physics of the I* and F* phases should be different in essential aspects from the physics of the C* phase, especially in situations related to polarization reversal and "surface stabilization".

**Bistability, domain walls and disclinations**

We see that the pictures of the initial experiments are very confusing. How about the initial model of bistable switching, is that more clear? In the initial papers, Clark-Lagerwall presented the model where there are two symmetric, bistable states of uniform orientation of the polarization vector, up or down orthogonal to the glass plates. This bistability is controlled by symmetric, bistable boundary conditions on both the boundary surfaces, where the molecules are forced to be parallel to the surface, but are free to rotate in the surface. In the recent articles, Lagerwall discusses also other possible configurations of liquid crystal cells, but does not explicitly discuss if the initial model was adequate for the initial measurements. Instead the model is denoted *"the basic geometry of a surface-stabilized ferroelectric liquid crystal"* (Lagerwall 1999) (p. 169) and used as a starting point for Lagerwall's claims that a SSFLC is a ferroelectric something, and the nice symmetry properties of the model are emphasized. If we look very closely, we might find some reservations:

*"Operation in the I\* phase has the advantage of a higher degree of bistability, assumed to be due to the ordering within the layers. In the C\* phase the stability is more sensitive to the surface treatment than in the I\* phase… The first (comment) is that it is not well understood why different compounds are strikingly and qualitatively different in their electro-optic behaviour (except for the obvious and measurable difference in macroscopic polarization) but very probably this has to be referred to their different interaction with surfaces, which is one of the least understood relevant phenomena in applied liquid-crystal science."* (Clark, Lagerwall et al. 1985)

*"Thus SSFLC cells are highly complex, and a complete quantitative understanding of these characteristics is not available for any cell. In many ways the development of devices has been carried out with and has been hampered by inadequate understanding of the operative cell physics."* (Lagerwall, Clark et al. 1989)

*"After unsuccessful attempts to explain all of the observations, we settled on a model, described next, which worked approximately for most of the cells we had studied and which described completely what we found in some cells."* (Clark and Lagerwall 1991) (p. 27)

*"Strangely enough, this simple geometry, which corresponds to the original concepts of (Clark and Lagerwall 1980a; Clark and Lagerwall 1980c; Clark and Lagerwall 1981), has only been realized in recent times, thanks to a devoted synthetic effort by chemists over the last 15 years. Before these materials are coming into use (not yet commercial) the layer structure is much more complicated."* (Lagerwall 1999), (hidden in the figure text of Fig. 61, p. 169).

Thus, it maybe was not only our incompetence that made it difficult for us, Lagerwall's graduate students, to reproduce the initial experiments… (Flatischler, Skarp et al. 1985). One fact, that we found, was that shearing of thin cells in general did not yield bistable cells. Instead we obtained mostly monostable cell, with an internal geometry resembling the proposed geometry in antiferroelectric cells with V-shaped switching (Dahl, Lagerwall et al. 1987). Lagerwall comments this paper in his book: *"However, the chosen form of the elastic torque means that it cannot describe the properties of a bistable cell"*, and instructed a later graduate student to rewrite the paper, without referring to it, removing the chosen form of the elastic torque(Hermann 1997), chapter 2.3. The rewritten paper of course still does not describe the properties of a bistable cell. When Lagerwall describes the basic geometry, it is still something real and existing, even "common" (Lagerwall 1996) demonstrated by the early experiments, and giving unproblematic "symmetric bistability".

"Symmetric bistability" is a stronger concept than "bistability". The symmetry implies that there is no reason for one of the states to decay to the other state, ever. The paper (Clark, Handschy et al. 1983) discusses a relaxation time of order 1 ms in the C* phase and essentially permanent memory in the F* phase, while the first paper (Clark and Lagerwall 1980a) talks about stability over several hours, not specifying the phase. Probably Lagerwall uses the word "symmetric" in a less stringent way.

Evidently the symmetry principles were not strong enough to ensure the success of the basic geometry. But did the basic geometry work for the initial experiments? There are strong reasons to believe that it was not so.

Besides the layer structure, there are some further problems with the basic geometry. The first thing is that the basic geometry is said to give two equivalent stable states and symmetric bistability. But it is also possible to obtain at least 4 additional states, where one of the surfaces is switched by the electric field, while the other is not, and where we have a deformed "splayed" state of the liquid crystal in between (3-state switching). The polarization has two easy ways of turning from pointing up at one surface to pointing down at the other. The stability of these "intermediate" states depends on several factors, but in general at least one of these states should show up intermittent during low voltage switching, at least as an occasional broadening of the Block or Néel walls, which are said to separate the ferroelectric domains. The domain walls are sharp, and are said to be deftly moved back and forth by varying the voltage.

We have then the problem with the polarity of the surfaces, which is discussed by Lagerwall in several places, but not as relevant to the basic model.

*"Our model incorporated what was a radical proposal, namely, that the stabilization of multiple states involved switchable interfaces, i.e., surfaces which could stably or metastably bind **n** in more than one orientation which could be switched. These have since been unambiguously demonstrated (Xue, Clark et al. 1988)"* (Clark and Lagerwall 1991) (p. 27)

What Xue really shows is that it may be possible to switch the surface by enough strong voltage (30 V in one direction, a few volts in the other), but the switching is very asymmetric in voltage, and the process is first order, with a large hysteresis. Xue manages to achieve switching of one boundary at a time, and there is no indication of any influence of the liquid crystal elasticity in the process. This kind of surface switching will not result in sharp domain walls between up and down domains, where the walls deftly move back and forth. Surface attached walls do not behave in that way. Clark and Lagerwall may have managed to find small regions with chevron bistability in their samples in the C* phase, and they may have seen the movement of disclination lines at the chevron interface.

There was a further observation made, supporting this, present in the video recordings from 1979. When observing the movements of the "domain walls" in the smectic C* phase by varying the voltage, in cell areas similar to fig 2b in (Clark and Lagerwall 1980a), the observed transmission in each of the states was strongly dependent of the weak voltage applied. Noel Clark, who did all these video recordings, and probably all experimental observations, described the transitions as *"continuous – domain wall – continuous"*. This is not at all consistent with the basic model, but consistent with the observation of chevron bistability. Zig-zag ("defect") lines were also seen in the sample, and Clark noticed that *"The domain walls can never cross the defect lines"*. The observed switching in the F* phase, and eventually also in the C* phase, was 4-state, which I interpret as a chevron transition near zero volt, and two surface transitions, one at positive voltage at one surface and the other at negative voltage at the other.

To differentiate between different types of bistability, it is important to investigate the domain boundaries. The basic geometry, with bistable switching of the boundary condition on both surfaces, implicates surface-attached walls, with or without an attached internal volume disclination line. The interface between the wall and the surface may be considered as a surface disclination line, so the basic model implicate two surface disclination lines, one at each surface. Chevron bistability implicates borders controlled by a disclination line at the internal chevron surface. Bistability with fixed boundary condition(Dahl and Lagerwall 1984) implicates an internal volume disclination line at the domain border. In the experiment by Xue (Xue, Clark et al. 1988), the surface domains are limited by single surface disclination lines. It is interesting to see what Lagerwall writes about bistability and the domain boundaries.

In their first paper (Clark and Lagerwall 1980a) Clark and Lagerwall say that the presence of a helix in their samples, with the defined boundary conditions, would require a disclinated structure, while their UP and DOWN domains are separated by well defined $\pi$ inversion walls. It do not seem to be evident to them that a helix structure can be seen as a striped structure with alternating UP and DOWN domains, separated by the same type of walls. (However, that fact is recognized in the Lagerwall book (Lagerwall 1999)) In the figure text of Fig. 1. a "disclination" inside the wall structure is discussed, but no disclination line is present in the figure itself. In the paper (Clark, Handschy et al. 1983), the boundaries are said to be sharp Bloch or Néel domain walls of approx 2 $\mu$m thickness, referring to pictures, which maybe show liquid crystals in the F* phase. To me these boundaries look very similar to the appearance of disclination lines in nematic liquid crystals. 1989 (Lagerwall, Clark et al. 1989), the word "disclination" is used both for the zig-zag wall in chevron samples and for a twisted cell structure, so it is not evident that the authors have a clear conception of the concept.

In the later articles, the basic geometry is presented as a matter of fact, but nothing really new is said about the physics of it: However, the following is said in a discussion about chevron bistability, which may or may not be accompanied by switching at the top and bottom surfaces:

*"However, bistability is always a surface effect, mediated by bounding surfaces, chevron surfaces or both"*. (Lagerwall 1999) (p. 310)

This is in clear contradiction to the configuration in Fig. 90 in the book, initially suggested by me.

**Implicit assumptions behind the basic model**

To evaluate the value of a physical model, it is often illuminating to study the explicit or implicit assumptions behind the model. For the basic model of the SSFLC structure, the following assumptions seem to have been made:

1. The substrate surface can be considered as an ideal, inert surface with symmetric, well-defined electric properties, interacting with an ideal liquid crystalline medium

2. The liquid crystal can be considered to be "stiff", so that the elasticity of the liquid crystal forces the both surfaces to switch at the same time. The surface switching is also a collective process, involving surface regions of at least micrometer size.

3. The reorientation between stable states is an "elastic" process, involving a potential function with two minima. This potential function is sketched in Figure 11 in (Lagerwall and Dahl 1984).

These assumptions are a bit problematic. Besides things already mentioned, a good inert surface will not give the smectic layer structure any support, and then it will be difficult to align the smectic layers and to keep a good orientation of aligned layers. The switching process also becomes degenerate, with transition from the UP to the DOWN position via both sides of the smectic cone, which could create a lot of defect lines during the transition. Moreover, in real cells there is very often a monolayer of adsorbed molecules at the surface, and this monolayer often seems to interfere with the orientation and reorientation of the liquid crystal. The symmetry of the surface will then also be dependent of the actual symmetry properties when this monolayer was created or reoriented last time. We can then adopt two different strategies: either we try to create an ideal, inert surface and that way eliminate the effects of the adsorbed monolayer, or we try to tailor the surface by molecular engineering to behave in the desired ways and enhance the mechanisms or maybe even create new mechanisms by attaching functional groups to the surface. The second approach opens many more doors. Thus we could formulate an alternative to point 1:

1. We should study the substrate surface on a molecular, stereo-chemical level and consider the electrochemical properties and the interaction at the molecular level with the free liquid crystal molecules.

The second point above touches my own obsession for studying the geometrical configuration inside the liquid crystal cells. The internal logic of the problems has forced me to also understand the importance of the boundary conditions. The basic idea of the SSFLC cell seems to be to unwind the helix by surface forces. This unwinding will occur essentially independently of the strength of the elastic forces, as long as they are weaker than the surface forces, since we have a competition between different geometrical configurations, where elastic volume forces are balanced by other elastic volume forces. So the unwinding of the helix does not imply that the liquid crystal can be considered as stiff (*"The unwound state thus corresponds to a smectic C medium where the infinite, continuous director degeneration has been lifted to only two-fold by the surface action…"*, (Lagerwall and Dahl 1984)). On the contrary, the elastic forces often are weak, and cannot be expected to be able to conquer over e.g. polar boundary conditions, unless electric forces help them. Thus, an alternative to point 2 would be:

2) We must consider the geometrical configuration inside the liquid crystal cells, and the multitude of degrees of freedom associated with this configuration. The surface switching process, when occurring, should be considered as a special process, separate from the reorientation of the liquid crystal but interacting with it.

When we consider the cell as an object with a multitude of degrees of freedom, it is easy to see the inadequateness of the picture of the potential function with two minima. In a multidimensional system, there very often exist tunnels through or passages around elastic energy barriers. However, traversing these tunnels or passages means scattering of energy in an unrecoverable way to many different degrees of freedom. This is usually denoted friction or viscosity. In our liquid crystal case, the passage of a disclination line might represent one tunnel in the energy barrier. Thus the alternative to point 3 should be:

3) The reorientation between states is a viscous process, involving dissipative processes both at reorientations at the surface and inside the liquid crystal. The actual path followed depends on the energy available. The detailed nature of the switching process and the nature of the borders between differently switched areas reveals the detailed physics of the system.

In the Lagerwall book, we cannot find very little about the surface physics of different materials: polyimide, evaporated $SiO_2$ and special rubbing techniques are mentioned in connection with some devices, but the surface properties are evidently seen as a problem for applications, not as a relevant problem for the physics. There are a lot of questions that never are asked. To what extend do the molecules stick to the surface? How will they be aligned then? What is the characteristic time for replacement or reorientation of these molecules? If we observe a reorientation during switching, is it the applied electric field that acts directly on the molecules, or is the electrical current or influence from the free liquid crystal molecules that cause the reorientation? Are molecules adsorbed in a polar or non-polar way? Does the nature of the surface influence the size of the polar surface interaction? If we have an elastic potential for the reorientation of a bistable liquid crystal, how high is the energy barrier? Is that barrier height dependent on the chemical nature of the surface? What is the role of the topography of the surface?

For a liquid crystal physicist it is natural to think about viscous forces as something that is proportional to the speed of change. Weak forces are then always accompanied by slow changes. But in the school physics we learnt about static friction, which could keep a box in complete rest on a slope, in spite of a finite gravitational force. Is there something like that in the field of liquid crystals? Yes, one example is the wide definition of SSFLC, where the surface forces are assumed to unwind the helix in thin cells. In spite of the weak twisting force, the helix will remain unwound forever. To my opinion, we have a similar case if we want to change the surface boundary conditions for a liquid crystal. The energy spent will be proportional to the size of the area, but not necessarily related to the speed of change. This energy can never be regained since the surplus will be dissipated as heat. Thus we have exactly the case of static friction. Thus a liquid crystal may be kept in one state in a bistable situation by static friction, as an alternative or complement to an elastic barrier. It is reasonable that this should be the case in <u>all</u> the situations where we have surface-controlled bistability at solid surfaces. The physics at internal surfaces, as for example the chevron surface, may be different. I have tried to discuss this question with Lagerwall, but he found this idea ridiculous and refused to discuss in a serious way.

The physics of the geometrical configuration inside the cells is discussed more in the book by Lagerwall, but it is characteristic that he prefers a nematic description of the elasticity theory for the smectic C* phase, in a similar way to the ignorance of the difference between the C* and the F*/I* phases in the initial experiments. It is also characteristic that he states (on page 316) that the splay in polarization is an inherent, spontaneous property of the smectic C* phase, and not just determined by the surface properties. This distinction is relevant for the discussion of the stability of the smectic C* phase and for the determination of the physics of disclinations, but not in the context of Lagerwall about the physics of thick pitch-compensated samples. Moreover, this effect can also be compensated, in a similar way as the pitch-compensation.

One way Lagerwall uses to handle question about the geometrical configuration is to state that there should be a characteristic length involved, and beneath this length the elastic forces should be able to cope with any external forces or polar boundary conditions. It is unclear to me if Lagerwall then consider the liquid crystal as crystalline in the solid-state sense, with a reduced number of degrees-of-freedom, or if he still consider it as an elastic medium, that just happens to be approximately aligned. However, the whole concept of characteristic length is of no meaning if it cannot be made plausible that this length is larger than the molecular dimensions.

Another indication of lack of interest for the geometrical configuration is the chapter about dielectric spectroscopy. By this spectroscopy one can both obtain information about the geometrical configuration of the liquid crystal and about the material parameters of the liquid crystal "as such". But the whole chapter is devoted only to material parameter determination, using configurations where the geometrical configuration is assumed to be known. Why is it so, do we really know everything that is relevant about the geometrical configurations, or is such knowledge not relevant?

Also the detailed switching process and the physics of walls and disclinations is treated superficially.

**Who invented SSFLC?**

From the patent application: *"The orientation in the sample will be separated by domain wall illustrated by dashed lines 144, the structure and width of which will be determined by the energy required to alter the tilt angle, the layer compression energy, and, most importantly, the surface energy and the bend and twist energies. In 1.5 ms thick DOBAMBC and HOBACPC samples, the domain walls are less than 1 ms thick. The application of an electric field will alter the orientation of molecules in the domain wall region to expand the favoured orientation.*

*Domain walls interact with defects in the layer structure, scratches and pits in the surface, and particulate impurities in the bulk and on the surface. These interactions have the effects of maintaining the positions of the wall once moved to a particular place by an applied electric field. As a result, the domains exhibit hysteresis, which gives threshold behaviour and*

*bistability. Bistability results since upon completion of a short application of an electric field applied to favour, for example, orientation 138, the domain wall will be stably pinned so that the sample will remain in orientation 138. Similarly, the same holds for the opposite polarity field with orientation 136. Threshold behaviour results because once the walls become bound, it takes an applied force larger than some critical value to dislodge them."* (Clark and Lagerwall 1980b).

Compare this to the following citation from the book, page 70-71:

*"The concept of surface stabilization has frequently been misunderstood or misinterpreted and various descriptions have been given which are physically incorrect (Fukuda, Ouchi et al. 1989), (Surguy 1993), (Surguy 1998). The most common mistake is to mix it up with "hysteresis", which is a word that can have a very wide usage, especially in various kinds of pinning effects, but should not be used in that sense for equilibrium states. Here reference (Clark and Lagerwall 1980a) has repeatedly been cited as a confirmation of hysteresis or even bistability and macroscopic polarization predicted in (Meyer 1977) as a result of suppressing the helix, without caring for what that article really contains (Fukuda, Ouchi et al. 1989), (Surguy 1993), (Surguy 1998). It might therefore be of interest to cite what actually is said in this review from 1977, and even do it in its context. What Meyer says about this matter is (page 39):*

'Another aspect of the ferroelectric response to an applied field is hysteresis. In crystalline ferroelectrics, in which there may be only a few easy axes for polarization, domain walls can have a high energy, due mainly to crystalline anisotropy. The pinning of domain walls, or the difficulty of nucleating new ones, is a major cause of hysteresis effects. In a single crystal chiral smectic C, there is no easy axis for polarization in the smectic layers, and thus there are no spontaneous domains; only line defects are allowed. Therefore in principle hysteresis is not possible. However in polydomain samples, or very thin ones contained between surfaces with strong alignment anchoring, there can be pinning effects which produce at least partial hysteresis. This is easily observed in experiments involving unwinding of the helix. When E is reduced below $E_c$, the helix reappears non-uniformly by the generation of discrete twist walls which nucleate on defects. At E = 0, the equilibrium pitch may not be achieved after such an experiment.'

*As we see, Meyer is referring to the helix unwinding by an electric field and in a different geometry, the layers being parallel to the glass plates, and the reappearing of the helix associated with hysteresis. He can observe this by the motion, reappearance, and coalescence of unwinding disclination lines. These effects have nothing to do with either the bistability or the hysteresis in helix-free bookshelf cells in which the unwinding lines are permanently eliminated. That symmetric bistability (the concept of which did not exist in any form prior to 1980) appears in this case, has nothing to do with whether the alignment anchoring is strong or weak, it is even essentially independent of what the surface conditions actually are. Therefore (Clark and Lagerwall 1980a) is not a proposal to use a kind of pinning hysteresis (Fukuda, Ouchi et al. 1989; Surguy 1993), but rather a demonstration of bistable switching together with an explanation of its origin, describing for the first time the significance of surface stabilization. As stated above, the expression SSFLC itself was not coined until 1983 in reference (Clark, Handschy et al. 1983)."*

First we can see that the citation from the patent application above seems to be inspired by the citation from Meyer: I know that Lagerwall has read that article by Meyer very carefully, underlined many passages and annotated in the margin. Then we also see that Lagerwall´s interpretation of the citation from Meyer simply is not plausible. If the smectic layers are parallel to the glass plates, we do not have a polydomain sample. And what does strong alignment anchoring mean for parallel layers? And how can strong alignment anchoring stop the reappearance of the helix in this case? For parallel smectic layers the reapperance of the helix usually occur by the motion of disclination lines inside the volume of the liquid crystal, not hindered by surface defects. But Meyer uses the concept "twist walls", which makes sense only if the smectic layers are not parallel to the glass plates. In Meyer's article he discusses both measurements when the smectic layers are parallel to the glass plates, and when then are normal to the glass plates, so he was aware of both possibilities. He explicitly mentions hysteresis effects in helix pitch measurements when the layers are normal to the glass plates. We may also cite further from the article by Meyer:

*"The potential application of the electro-optical effects in the ferroelectric phase have only begun to be explored. Especially interesting are the possible switching effects in very thin layers between conducting glass plates, in which surface pinning effects may be utilized to achieve an electro-optical memory."*

So the real inventor of the SSFLC cell is R.B. Meyer, maybe with support from Ph. Martinot-Lagarde, R. Duke, and G. Durand:

*"The sample thickness (d = 150 **m**m) is chosen to avoid the helixquenching by the walls. (We have indeed observed (Duke, Durand et al.) this quenching when the thickness is about the pitch (d = 20 **m**m)." …"Nevertheless the dispersion of the field value needed to put out the lines suggests that in this geometry this field is not only the critical field to unwind the helix but also the field necessary to destroy a wall anchoring."* (Martinot-Lagarde 1976) Thus Martinot-Lagarde suggests polar electrically-controlled, multistate boundary conditions, and in the paper he also mentions hysteresis effects connected to field reversal.

Also L.J. Yu, H. Lee, C.S. Bak, and M.M. Labes have reported about memory states with macroscopic dipole moment in untwisted smectic-C* thin (6.3 μm) samples, with planes perpendicular to the glass plates (Yu, Lee et al. 1976), and they also confirm that R.B Meyer had seen long-lived, untwisted states.

We should also mention here that Clark and Lagerwall in their first paper(Clark and Lagerwall 1980a) states that M. Brunet(Brunet and Williams 1978) has demonstrated suppression of the helix in thin cells to form a unique stable SC monodomain, thus she has demonstrated a SSFLC according to the wide definition. She had suggested "a stronger boundary condition". Remember then again what Lagerwall states in his book:

*"That symmetric bistability (the concept of which did not exist in any form prior to 1980) appears in this case, has nothing to do with whether the alignment anchoring is strong or weak, it is even essentially independent of what the surface conditions actually are. "*

What did Clark and Lagerwall really invent or see?

    Ferroelectric liquid crystals: No.

    SSFLC according to the wide definition: No.

    SSFLC according to the narrow: Maybe in the F* or I* phase.

    Fast switching in the C* phase: Yes probably.

    Bistability in the C* phase: Probably of chevron type.

    Working model for SSFLC: Naïve.

We see that all the ingredients were already there at the time of the Clark-Lagerwall discovery: the chiral smectic-C phase, the orientation of the smectic layers orthogonally to the substrate to facilitate coupling to an applied electric field, the helix quenching by surface forces, and electrically controlled multistate surfaces providing hysteresis and memory states to the liquid crystal. There were a number of loose dangling ends, demanding someone that was able to tie up the sack. Voilá, Clark-Lagerwall found the Columbus´ egg, and the SSFLC device was invented. But, and that is a big "but", reality was not so simple and obedient as it ought to be to the great physicists. The chemical substances used were dirty, unstable and unreliable, and the orientation in the smectic-C phase most of all resembled porridge. I think

Noel Clark was realistic, and saw the "discovery" as a starting point to a new and unfinished research field, while Lagerwall has been carried away and astray by the fame and public attention, and tries to transform the immature SSFLC device specification to a dogma about ferroelectricity. Maybe the biggest miracle was that they managed to patent the idea.

**The rest of the book**

What about the rest of the book? I am not an expert in all of the details that are discussed in the book, but in the areas where I have some insight, I see that Lagerwall is not entirely wrong. But he is not entirely right either: the description is just shallow and unreliable, and biased by private politics, promoting himself and his friends. This means that several young researchers, fresh in the field, will have to struggle and waste time to sort out more or less trivial errors, made by the guru of the field. Some examples: Lagerwall states that the helical pitch in the C* phase should diverge at the transition to the A phase ("this is a priori natural, as no helix is allowed in the A* phase"). This statement, also included in *Handbook of Liquid Crystals(Demus, Goodby et al. 1998)*, is supported by a simple mathematical error, where he erroneously has used the expression for the cholesteric twist to calculate the twist in the smectic C* phase, ignoring the tilt of the molecules relative to the twist planes. Maybe this is a trivial error, but it is severe for someone who wants to be the expert of the field. Lagerwall has had discussions with his graduate students about this point. I also strongly remember a discussion I had with Lagerwall about the necessity for a pitch divergence. I made the analogy with a metal spiral spring or coil, which we dissolve in an acid. The helix does not need to unwind; it can vanish with a finite pitch. In the smectic A phase, there might even be chiral correlations between the tilt fluctuations in adjacent layers, and from these correlations we might define a finite pitch length, even if the helix is absent, almost like a grin without a cat(Caroll 1865). Lagerwall found the coil analogy ridiculous. I consider "ridicule" to be a social concept, not really relevant or adequate in a serious scientific discussion.

Another example is the chapter about dielectric properties. Lagerwall has managed to make systematic errors in the derivation of the dielectric tensors, and the errors also have propagated to or from the chapter about viscosity. To check the dielectric tensors, one could easily calculate the direction of the eigenvectors, and one of them should be aligned with the nematic director. Lagerwall has also the papers of his graduate students available, and he should have checked with those, but he evidently has not done that. As an example, in Eq. (375) I find three independent errors, where only one could be attributed to a printing error, propagating from *Handbook of Liquid Crystals(Demus, Goodby et al. 1998)*.

As a last example of private politics, involving myself: In the index, there is something denoted the "Lagerwall reference frame", see Fig. H. Lagerwall has thus considered this name to be appropriate for the concept. When I started looking at the elasticity theory of the smectic C phase, the only paper available was that by the Orsay group(Orsay Group on Liquid Crystals 1971), where the deformations where described by small rotations. I wanted a description that was more similar to the nematic description, and started, on my own initiative and without the involvement of Lagerwall, to reformulate the elasticity theory. As supervisor, he was included as co-author. In the nematic phase, we have a unit vector, the director, available for the description. In the smectic C phase, we can, from the symmetry properties, in every point define an orthogonal base of vectors, and the elasticity theory can be described in terms of the spatial derivatives of this base. In his book, Lagerwall insists that it is more useful to have a description in terms of the nematic director (not solidly based on the symmetry properties), but he anyway wants to name the reference frame after himself. In a recent report to the new Ethical committee of the Swedish Research Council, he accuses a colleague for dishonest use of the results of students and co-workers.

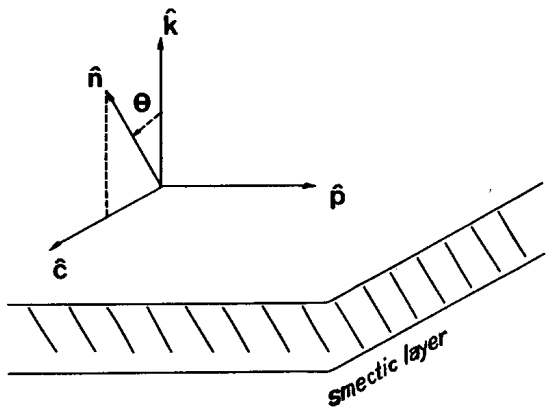 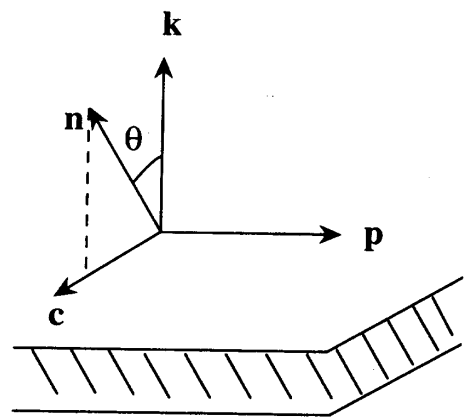

Fig H: My reference frame (Dahl and Lagerwall 1984)

Lagerwall reference frame 1999 (Lagerwall 1999)